\documentclass[]{aa}
\usepackage{graphicx}
\usepackage{natbib}  
\bibpunct{(}{)}{;}{a}{}{,}
\usepackage{txfonts}
\usepackage[]{hyperref}

\defcitealias{PALTA2}{Paper~I}
\usepackage{amstext}

\begin{document}

\title{Astrometric planet search around southern ultracool dwarfs}
\subtitle{IV. Relative motion of the FORS2/VLT CCD chips\thanks{\textit{Based on observations made with ESO telescopes at the La Silla Paranal Observatory under programme IDs 086.C-0680, 087.C-0567, 088.C-0679, 089.C-0397, 090.C-0786, 091.C-0083, 092.C-0202, 596.C-0075, and 0103.C-0428.}}}

\author{P. F. Lazorenko\inst{1}
                \and J. Sahlmann\inst{2}}

\institute{Main Astronomical Observatory, National Academy of Sciences of the Ukraine, Zabolotnogo 27, 03680 Kyiv, Ukraine\\
                \email{laz@mao.kiev.ua}                         
                \and
                Space Telescope Science Institute, 3700 San Martin Drive, Baltimore, MD 21218, USA}

\date{Received ; accepted  }
\abstract{
We present an investigation of the stability of the two chips in the FORS2 camera CCD mosaic on the basis of astrometric observations of stars in 20 sky fields, some of which were monitored for four to seven years. We detected a smooth relative shear motion of the chips along their dividing line that is well approximated by a cubic function of time with an amplitude that reaches $\sim$0.3 pixels (px) or $\sim$38 mas over seven years. In a single case, we detected a step change of $\sim$0.06~px that occurred within four days. In the orthogonal direction that corresponds to the separation between the chips,  the motion is a factor of 5--10  smaller.
This chip instability in the camera significantly reduces the astrometric precision when the reduction uses reference stars located in both chips, and the effect is not accounted for explicitly.  We found that the instability introduces a bias in stellar positions with an amplitude that increases with the observation time span. When our reduction methods and FORS2 images are used, it affects stellar positions like an excess random noise with an RMS of $\sim$0.5~mas for a time span of three to seven years when left uncorrected. We demonstrate that an additional calibration step can adequately mitigate this and restore an astrometric accuracy of 0.12~mas, which is essential to achieve the goals of our planet-search program. These results indicate that similar instabilities could critically affect the astrometric performance of other large ground-based telescopes and extremely large telescopes that are equipped with large-format multi-chip detectors if no precautions are taken.}

\maketitle

\section{Introduction.}{\label{int}}

Astrometry is a powerful technique for detecting companions to the nearby ultracool dwarfs by monitoring the induced reflex motion, provided that a measurement precision better than 1 mas can be achieved. For ground-based observations, this requires the use of 8--10~m telescopes and good observing conditions. For targets near the Galactic plane, where a large number of reference stars is available, and  0.4\arcsec\ seeing,  the initial  estimates of \citet{LazLaz} indicated an achievable astrometric accuracy of 0.01--0.06~mas for faint stars.  With the Very Large Telescope (VLT) camera FORS2 \citep{FORS} we demonstrated a single-epoch astrometric accuracy close to 0.1~mas  for 15--20~mag stars and 0.6\arcsec\  atmospheric seeing \citep{Lazorenko2009, PALTA1, PALTA2, gaia}. This accuracy is  sufficient to detect and characterize the orbits of Jupiter-mass companions to nearby ultracool dwarfs.  With this purpose,  we initiated an astrometric planet search around 20 ultracool  dwarfs in 2011 \citep{Sahlmann2015DE0823, Sahlmann_palta3}.

The further accuracy improvement beyond 0.1~mas, needed for the detection of Neptune- and lower-mass companions to ultracool dwarfs,  is restrained  by  fundamental limits on the precision of photocenter determination, the atmospheric image motion caused by turbulence in the upper atmospheric layers, and by systematic errors of instrumental origin.  It is expected that in the near future the achievable astrometric accuracy will be  improved to 0.01--0.05~mas on 30~m class telescopes such as the European Extremely Large Telescope (ELT), primarily  due to the use of adaptive optics, better signal-to-noise ratio in the stellar images, and general performance improvements of the cameras \citep{MICADO2010,MNRAS_ELT,SPIE_ELT}.

The CCD detectors in most  cameras of large telescopes are formed as mosaics of CCD chips  allowing for a wide field of view (FoV). However, this structure of  detectors potentially affects   geometric stability of the compound image when the mounting of the CCDs  is not sufficiently rigid. The first evidence of the  displacements in the FORS2 CCD mosaic were detected by   \citet[hereafter \citetalias{PALTA2}]{PALTA2} from the analysis of two years of data. At that time, we discovered a  relative chip motion (RCM) in the two-chip CCD mosaic with an  amplitude of   0.001--0.010~px, or 0.1--1~mas.  In one peculiar case, we detected a shift of nearly 0.02--0.05~px or  3--6~mas. These amplitudes were small but exceeded the astrometric accuracy of FORS2 by a factor of 10. 

Relative chip motion was observed for cameras on board the Hubble Space Telescope \citep[][Sect. 6]{Anderson:2003aa}\footnote{See also \url{http://www.stsci.edu/itt/review/dhb_2010/WFPC2/wfpc2_ch55.html\#1961580}.} with an amplitude of up to 3~px=120~mas over $\sim$10 years.
More recently, \citet{dark} analyzed the stability of the CCD chip array of the Dark Energy Camera (DECam)  and discovered surprisingly large RCM with an amplitude of 0.4~px=100~mas occurring over only 2 weeks.

In this paper, we use newly obtained  series of FORS2 observations to characterize the instability of its detector chips. We also predict how these instrumental errors can affect the astrometric performance of FORS2 and other telescopes.

\section{Observational data}{\label{obs}}
This study is based on observations of $N_{\rm dw}=20$ ultracool  dwarfs  obtained with the FORS2 camera  of the VLT. The program of observations, reduction methods,  and the first results obtained from observations in 2010--2013 are described in \citet{PALTA1} and \citetalias{PALTA2}. We monitored all targets during two visibility periods, but for targets with suspected orbital signals, we followed up with additional observations. As a consequence, six fields were observed over one to one and a half years, seven fields were observed for three years, and the seven remaining fields were monitored over six to seven years. We refer to the corresponding data as the short-, medium-, and long-duration datasets.

{ The FORS2 detector has an exchangeable blue (E2V) and red (MIT) sensitive detector pair. We used FORS2 only in its  standard configuration with an MIT detector. This detector is a mosaic  of two CCD chips.} With the 2$\times$2 binned read-out mode we used,  they are 2k$\times$1k pixel arrays with pixel sizes of 30~$\mu$m. The upper chip 1 and the lower chip 2  are mounted with a small gap of $\sim$480 $\mu$m ($\sim$16 binned pixels).  Observations were obtained with a high-resolution collimator that produces an angular pixel scale of 126.1~mas/pixel. Every observation `epoch` consists of a series of 30--100 images. For the different sky fields we obtained  $N_e=10 \ldots 23$ epochs of observations.

In every image, we measured  500--1500 star photocenters  $x$, $y$ of $i=1 \ldots N$ unsaturated reference stars. The following reduction proceeded in a coordinate system  whose axes $X$, $Y$ are oriented along RA, Dec. For convenience, the row $y=1000$~px was aligned  with the upper edge of the gap between the chips (a lower edge of the upper chip 1), thus  $y<1000$~px identifies stars in the bottom chip 2, and  $y>1000$~px corresponds to the upper chip 1.

\section{Basic astrometric reduction}{\label{red}}
The observations are reduced {with differential astrometric techniques} as previously reported (\citealt{Lazorenko2009}, \citetalias{PALTA2}), where we use the effective filtration of the atmospheric image motion by the large ground-based telescope \citep{LazLaz}. This technique  requires a circular reference field, and the best performance is achieved at  its center.  For this reason, the reference field  is centered on  the target object. We take into consideration that the  atmospheric image motion variance increases with the reference field size $R$ while the reference frame noise decreases with $R$. Therefore, there is some optimal radius $R_{\rm opt}$ at which the sum of these error components is smallest. Its value is determined numerically and is 400--800~px (1--2\arcmin \, for FORS2), which is typical for the sky fields near the Galactic plane. 

{ A list of the variables  that are most frequently used in the paper along with a  short description is given in Table~\ref{var}.}
\begin{table}[tbh]
\caption [] {List of frequently used variables and functions}
{\small
\begin{tabular}{@{}l@{\quad}l@{}}
\hline
\hline
quantity & definition               \rule{0pt}{11pt}\\
\hline
$i$ / $m$ / $e$ / $z$ & indexes for the star, frame, epoch, and sky field number\\ 
$x_{i,m}$, $y_{i,m}$    & measured star $i$ position in the frame $m$         \\
$x_{i,0}$, $y_{i,0}$    & star $i$ position in the reference frame $m=0$       \\
$\Phi^x_m$, $\Phi^y_m$ & geometric field distortion \\
$\phi^x_m$, $\phi^y_m$ & displacement due to the proper motion, parallax, \\
& and differential chromatic refraction  (DCR effect)\\
$\xi_{i,s}$  & star $i$ model astrometric parameter, $s=1 \ldots 7$ \\
$V^x_{i,m}$, $V^y_{i,m}$ & displacement of the star in frame $m$ relative  to frame\\
&  $m=0$,  corrected for the  geometric field distortion \\
$V^x_{z,e}$, $V^y_{z,e}$ & the same, averaged over some star subsample in the \\
& sky field $z$ at epoch $e$ \\
$\Delta^x_{i,e}$, $\Delta^y_{i,e}$ & positional residuals  for star $i$ in the epoch $e$\\
$g^x_{e}$, $g^y_{e}$ & systematic difference in the { epoch positional residuals} \\
& of chips 2 and 1 at the division line between chips\\
$H^x_z(t_e)$, $H^x_z(t_e)$ & the smooth component of the RCM in field $z$,  epoch $e$\\
$\delta H^x_{z,e}$, $\delta H^y_{z,e}$ & irregular component of the RCM  in field $z$,  epoch $e$\\
\hline
\end{tabular}
\label{var}
}
\end{table}

\subsection{Astrometric model}{\label{astr_mod}}
All measured stars (except for the target object) are used as references to provide the best information on the geometric distortion $\Phi(x,y)$ of the field caused both by the instability of the telescope optics and its adjustment, and by the image motion. { Its spatial power density is represented by an infinite expansion into series of even integer powers $k=2,4,...$ in the domain of the spatial frequencies, where $k$ is the modal index, and the mode amplitude decreases rapidly with increasing $k$. Over the CCD space ($x,y$),  the image motion causes deformations of the frame of reference stars  that are modeled by an expansion into a series  of basic functions $f_{i,1}=1$, $f_{i,2}=x_i$, $f_{i,3}=y_i$, $f_{i,4}=x_i^2 \ldots$ which are full bivariate polynomials of $x$, $y$ and whose number $N'=1 \ldots k(k+2)/8$ depends on  $k$. The use of functions $\Phi(x,y)$ in the reduction model removes all first $1 \dots k/2-1$ modes of the image motion spectrum. Usually, $k$ is  between  4 and  16, and for this study,  we set $k=10,$ which leads to the most stable solution. The  function $\Phi^x_m( x, y)=\sum_{w=1}^{N'} c^x_{w,m}f_{i,w}$ therefore includes $N' \times M$   coefficients $c^x_{w,m}$  that model geometric distortions to the fourth power of $x$, $y$ at the exposure  number $m=1\ldots M$. } Similarly, $\Phi^y_m( x, y)$ models geometric distortions along the $Y$ -axis through the coefficients $c^y_{w,m}$.

The displacement of some star $i$ photocenter (either the target or reference star)  from its initial position $x_{i,0}$, $y_{i,0}$ in the reference frame (exposure $m=0$) to  $x_{i,m}$, $y_{i,m}$  in exposure $m$ is caused by intrinsic (proper motion, parallax) and atmospheric (refraction) motion. We model it with $s=1\ldots S $ astrometric parameters $\xi_{i,s}$: the offsets to the positions $\Delta x_{0,i}$, $\Delta y_{0,i}$,  proper motion $\mu_{x,i}$,  $\mu_{y,i}$, relative parallax $\varpi$, and two atmospheric differential chromatic  parameters $\rho$ and $d$, that is,\ $S=7$. The corresponding displacement is $\phi_m^x(\xi, \nu^x)=\sum_{s=1}^S { \xi}_{i,s} \nu^x_{s,m}$  and $\phi_m^y(\xi, \nu^y)=\sum_{s=1}^S { \xi}_{i,s} \nu^y_{s,m}$  , where  ${\nu^x}$, ${\nu^y }$ are    $S\times M$  coefficients (time $t$, parallax factors $\Pi$, etc.) valid at the time of exposure $t_m$ for the  measurements along $X$ or $Y$. { In particular, $\phi_{m,i}^x(\xi, \nu^x)=\Delta x_{0,i} +\mu_{x,i} t_m + \varpi_{x,i} \Pi_{x,i,m} - \rho f1_{x,m} - d f2_{x,m}$ , where functions $f1$, $f2$ depend on zenith distance, parallax angle, temperature, and pressure \citep{PALTA1}. }

All $\xi$ values are determined as  free model parameters. Alternatively, the reduction model can be significantly  simplified if some $\xi$ values are taken from the external sources.    In particular,  {\it Gaia} DR2 \citep{DR2} provides proper motions with standard uncertainties of 0.08--0.16~mas~yr$^{-1}$ for  $G=$16--17~mag stars. For our targets of comparable magnitudes with a time base of about 1.4~yr, the accuracy of FORS2 proper motions is 0.10~mas~yr$^{-1}$  \citepalias{PALTA2}, which  improves to 0.01--0.03~mas~yr$^{-1}$ for the  objects  monitored over 5--7~yr. Because  the accuracy of differential FORS2  proper motions and parallaxes is currently better, we do not use DR2 as an external source of $\xi$ values at this time. However, we use Gaia DR2 as an astrometric flat-field calibrator, see Sect.\ \ref{gaia}. The differential chromatic refraction (DCR) parameters $\rho$,  $d$ can be constrained using stellar spectra \citep[e.g.,][]{Garcia}, however for field stars these are usually unavailable.

The measured displacement  of a star from the initial position $ {x_{0}}$, $ {y_{0}}$ is described by the sum of functions $\Phi(x,y)$ and $\phi(\xi,\nu)$, or
\begin{equation}
\begin{array}{ll}
\label{eq:eq1}
\sum\limits_{w=1}^{N'} c^x_{w,m}f_{i,w} + \sum_{s=1}^S { \xi}_{i,s} \nu^x_{s,m}   +H_m^x U(y)+ \delta^x_{i,m}  = x_{i,m} - x_{i,0}, \\
\sum\limits_{w=1}^{N'} c^y_{w,m}f_{i,w} + \sum_{s=1}^S { \xi}_{i,s} \nu^y_{s,m}    +H_m^y  U(y) + \delta^y_{i,m} =    y_{i,m} - y_{i,0}.\\
\end{array}
\end{equation}
This system is solved using reference stars $i \in \omega$ , where $\omega$ is the reference  space that includes either chip 1 only  or { stars of both chips within $R_{\rm opt}$ distance from the target}.   Equation (\ref{eq:eq1}) is solved with the weights $P_{i,m}$ assigned according to the measurement uncertainties $\sigma_{i,m}$ of the star $i$ in the frame $m$. In the context of this investigation, additional terms $H_m^x$, $H_m^y$  are included as the  potential displacement of chip 2  relative to chip 1 along the coordinate axes in exposure $m$.  In the primary astrometric reduction, the solution of Eq.\,(\ref{eq:eq1}) is derived assuming $H=0$, and later, during the second phase of the investigation,  the terms $H_m$ are extracted  by  analyzing  the model  residuals $\delta^x_{i,m}$,  $\delta^y_{i,m}$ and the peculiar properties of the $\xi$ parameters. The step function $U(y)$ 
\begin{equation}
\begin{array}{llr}
\label{eq:u}
 U(y)= 1   & \mathrm{if}  \hspace{2mm}   y<1000  &\mathrm{(lower\: chip\: 2)}\\
 U(y)= 0    & \mathrm{if}  \hspace{2mm}   y>1000  &\mathrm{(upper\: chip\: 1)}\\
\end{array}
\end{equation}
is introduced to limit the application of the $H$ function to one chip only.
The system Eq.\,(\ref{eq:eq1}) is   solved with $S \times N' $ conditions
\begin{equation}
\label{eq:eq2}
\sum_{i \in \omega^*}  \xi_{i,s} \bar {P}_{i}  f_{i,w}(x_i,y_i) =0, \quad s=1\ldots S,   \quad w=1\ldots N' 
\end{equation}
on the parameters $\xi$ in  the normalizing parameter space  $\omega^*$, which does not  necessarily coincide with $\omega$.  When $s$ is fixed,  Eq.\,(\ref{eq:eq2}) can be understood as the formal expression of the fact that  the least-squares  residuals (which are recognized here as $\xi_i$) are   orthogonal to  each of $w=1\ldots N'$  basic  functions $f_{w}$  of some conditional system of  $i=1 \ldots N$ measurements obtained with the weights $\bar {P}_i$. In this interpretation, the corresponding system of conditions is $\xi_{i,s}^{\mathrm{abs}}= \sum_{w=1}^{N'} c^*_{w,s} f_{i,w}$ , where  $\xi^{\mathrm{abs}}$  are the parameters given in some external (e.g., absolute) system. Consequently,  $\xi_{i,s} =  \xi_{i,s}^{\mathrm{abs}} - \sum_{w=1}^{N'} c^*_{w,s} f_{i,w}$ represents  the fit residuals  \citep{Lazorenko2009}. This establishes the relationships between  the system of relative  parameters $\xi$,  the absolute  parameters $\xi^{\mathrm{abs}}$, fit constants $c^*$, the weights $\bar {P}_i$, and  the functions $f_w$.  

One property of  the derived parameters $\xi$ is that the weighted sum of differential  proper motions  is  zero, that they do not change systematically across the FoV,  and that they do not correlate with $x$, $y$, $xy$ or any other function $f_w$, which is expressed by $\sum \mu_i \bar {P}_{i} = \sum \mu_i \bar {P}_{i} x_i = \sum \mu_i \bar {P}_{i} y_i =\ldots \sum \mu_i \bar {P}_{i}  f_{i,w} =0 $.  Equation (\ref{eq:eq2}) is applied with arbitrary weights $\bar {P}_i$ that define the system of relative  parameters $\xi$. In practice, we set $\bar {P}_i=1$ for sufficiently bright stars and $\bar {P}_i=0$ for  $\sim$$10\%$  for fainter reference  stars to exclude wildly scattered values $\xi$ of these stars. For the standard astrometric reduction, we usually set  $\omega^* =\omega$, but  in this investigation,  $\omega^*$ space can be expanded over $\omega$, see Table\,\ref{table}.  

The model Eq.\,(\ref{eq:eq1}) with constraints  Eq.\,(\ref{eq:eq2}) is solved iteratively as described in \citet{Lazorenko2009}, in turn  deriving $c_{w,m}$ within each single image $m$  at some fixed $\xi_{i,s}$. With these  $c_{w,m}$, we then update $\xi_{i,s}$ separately for each single reference star $i$. Here we omit the details of the reduction and note only that as soon as  iterations converge and the geometric distortions $\Phi_m$ of each frame $m$ are found, the $i$-th star's astrometric parameters $\xi_{i,s}$ are derived from the subsystem of equations
\begin{equation}
\label{eq:eq3}
\begin{array}{ll}
\sum_s \xi_{i,s} \nu^x_{s,m} + \delta^x_{i,m}  +H_m^x U(y) = V^x_{i,m}\\
\sum_s\xi_{i,s} \nu^y_{s,m}  + \delta^y_{i,m}  +H_m^y U(y) =V^y_{i,m}, \\
\end{array}
\end{equation}
where  $V^x_{i,m}=x_i - x_{i,0}- \Phi^x_m(x_i,y_i)$ and $V^y_{i,m}=y_i - y_{i,0}- \Phi^y_m(x_i,y_i)$ are the stellar displacements corrected for $\Phi_m$.  This system involves  $m=1\ldots M$  measurements of the star $i$ and is solved considering both $x$ and $y$ data simultaneously, which also yields the residuals $\delta^x_{i,m}$,  $\delta^y_{i,m}$. Because $\nu^x_{s,m}$,  $\nu^y_{s,m}$  are the basic functions of  the system Eq.\,(\ref{eq:eq3}) solved by the least-squares fit, the residuals $\delta^x_{i,m}$,  $\delta^y_{i,m}$  are   orthogonal  to these functions. In particular, they should have no linear dependence on time. Similarly, the residuals $\delta^x_{i,m}$,  $\delta^y_{i,m}$ , which are orthogonal to the basic functions  $f_{i,w}$ of  Eq.\,(\ref{eq:eq1}), should have no systematic  trend along $X$ or $Y$. The frame residuals of  reference stars, converted into the epoch residuals $\Delta^x_{i,e}$ and $\Delta^y_{i,e}$, can display  some correlation pattern that is caused by systematic errors, which provides useful information on these errors.  In particular, the  investigation of the epoch residuals derived under the assumption that $H(t)=0$ allows us to reveal the time-dependent relative chip motion.

\subsection{RF and CAL files}{\label{datasets}}
The principal results of the astrometric reduction  are the parameters $\xi_{i,s}$ and the epoch residuals $\Delta^x_{e}$, $\Delta^y_{e}$ derived for the star $i$ that is the target object.  In addition, we obtain the same quantities for every reference star within a circular reference field centered on the target.  These data in 20 sky fields are referred to as the reference frame (RF) files. The uncertainties of the RF dataset change across the field and are smallest at its center. At the outer boundaries, the astrometric accuracy degrades because the uncertainty of the $\Phi(x,y)$ determination increases. 

Stars in the RF files do not cover the full FoV for typical sizes of $R_{\rm opt}$. For calibration purposes we need to know the astrometric parameters and the epoch residuals of all stars in the full FoV.  We achieve this by repeating the basic reduction for each of $i=1\ldots N$ field stars, setting it as the target object. The results are saved in  the calibration (CAL) files that contain results for the single central star only. In CAL files the reference star data are disregarded because the precision is reduced.  The CAL files include most stars in the FoV and therefore contain about three times more stars than RF files.

\subsection{Response properties of the astrometric model}
\begin{figure}[tbh]
\resizebox{\hsize}{!}{\includegraphics* [width=\linewidth]{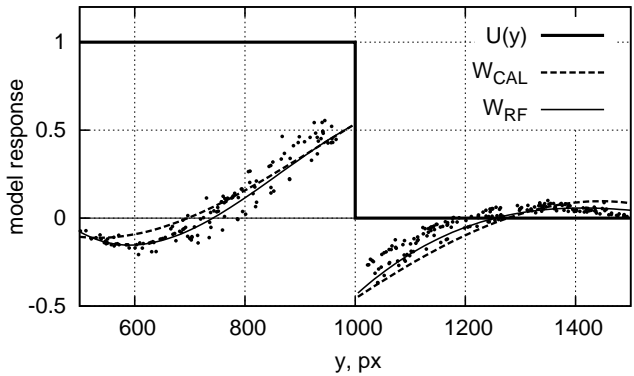}}
\caption{Simulation of the astrometric model response  to a step change $U(y)$  in the position of chip 2 for the example sky field of dw14 offset by  $73$~px from the detector center.  Dots show the response in the epoch residuals $\Delta^x_{i,e}$  of reference  stars, and the function $W_{RF}$   is the smoothed response in proper motions to the step change of proper motions of chip 2 stars. The function $W_{CAL}$  is the same,  but for the reference stars distributed evenly over FoV.} 
\label{responce}
\end{figure}

{\label{trans}}
The model defined by Eqs.\,(\ref{eq:eq1}--\ref{eq:eq2}) can be treated as a filtering system that absorbs some part of a signal at its entrance because of the least-squares fit of data. Therefore we should not expect that the chip displacement $H(t)$ that affects the measurements is fully reflected in the epoch residuals. We  demonstrate this with the help of reference stars around the brown dwarf \object{DENIS J125310.8-570924} (hereafter dw14), for which we obtained $N_e=23$ epochs over $\text{about seven}$ years. We introduced a step change $U(y)=1$ in the measurements  $x_{i,e}$ obtained on 17 February 2012  and performed our reduction process. Figure \ref{responce} shows that the computed residuals differ significantly from the introduced step change. The residuals have a discontinuity at $y=1000$~px with an offset amplitude of 1 between the chips. The scatter of individual $\Delta^x_{i}$ values relative to their systematic change as a function of $y$ is related to a weak dependence of $\Delta^x_{i}$ on $x$, and a small asymmetry relative to the detector center $y=1000$~px  is due to the location of the reference field center  $\sim$73~px off the gap between the chips. 

The solid curves in Fig.\,\ref{responce} correspond to the systematic change of the residuals if the reduction is made with $ {P}_{i,m}=1$. The restriction Eq.\,(\ref{eq:eq2}) leads to  $\xi$ parameters that are orthogonal to the functions $f$ with the weights $ \bar{P}_{i}=1$, that is, the response characteristics of  $\xi$ and $\Delta_{i}$ derived with $ {P}_{i,m}=1$ are exactly equivalent.  In this  interpretation, solid curves represent the model response $W_{RF}$  in $\xi$ (e.g., in  proper motions) to the step change of these parameters  for all chip 2 stars by $ U(y)$. The response $W_{RF}$ corresponds to RF files or to the proper motions of reference stars in a field centered at the target.

The dashed lines in Fig.\,\ref{responce} correspond to a reference field that is symmetric around the detector center   $y=1000$~px and with $ {P}_{i,m}=1$ set in Eq.\,(\ref{eq:eq1}). This option  approximately corresponds to the averaging of response functions over many circular reference fields with centers that cover most of the FoV, and we therefore used it as the  response function  $W_{CAL}$ for the CAL files. This function is nearly antisymmetric relative to $y=1000$~px, where it is equal to $\pm 0.5$. Below  we discuss the application of the response functions to the effects in the epoch residuals and in proper motions.
The above examples demonstrate that a step signal $U(y)$ is associated with a specific discontinuity pattern in the epoch residuals or proper motions.

\section{Observational evidence of the chip motion}
\begin{figure}[tbh]
\includegraphics[width=\linewidth]{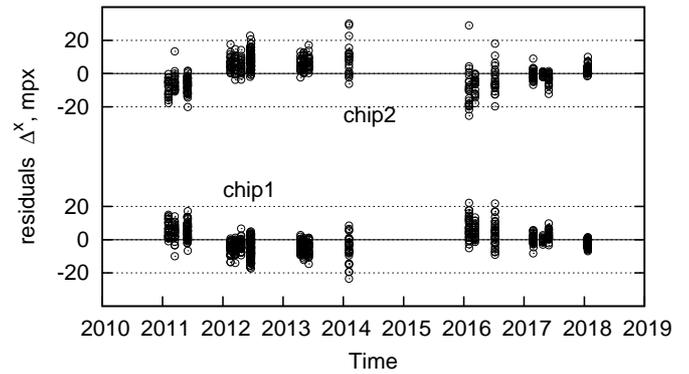}
\caption{Epoch positional residuals $\Delta^x_{e}$  for bright reference stars imaged close to the chip gap in the sky field of dw14. The residuals are shown  separately for stars in chip1 and chip 2 and exhibit an anticorrelated pattern.}
\label{dw14}
\end{figure}

{\label{s_mot3}}
First evidence of the FORS2 chip mosaic instability was found in the epoch residuals  of reference stars within the same reference field, but located in different chips \citepalias{PALTA2}. As an example,  Fig.\,\ref{dw14} shows the  residuals $\Delta^x_{e}$ in RA of stars in the sky field of dw14. The residuals  are shown separately for the stars imaged in each chip and for a few  bright stars only that were measured with a single-epoch uncertainty of  0.1--0.5~mas,  or about 1--3~millipixels (mpx). These residuals do not show a linear trend in time because the astrometric  model includes the proper motion as a model parameter. However,  a significant time-dependent variation with opposite sign in both chips is evident. 

In \citetalias[Sect.4.3.2]{PALTA2}, this  effect was explained  by the  relative chip displacement without deformation and rotation.  The  motion of chips along the CCD $X$ -axis for each epoch $e$ was described by the quantity $g_{x,e}$  equal to the systematic difference 
\begin{equation}
\label{eq:def_g}
g_{x,e}=  \Delta^x_{i,e}({y \to 1000^-}) - \Delta^x_{i,e}({y \to 1000^+}) \end{equation}
in the measured positional residuals of stars in chip 2 and chip 1 at the division line ($y=1000$~px). The motion $g_{y,e}$ along Dec ($Y$ -axis of CCD) was defined similarly as a difference of $\Delta^y_{i,e}$ values.

\begin{figure}[tbh]
\resizebox{\hsize}{!}{\includegraphics*[width=\linewidth]{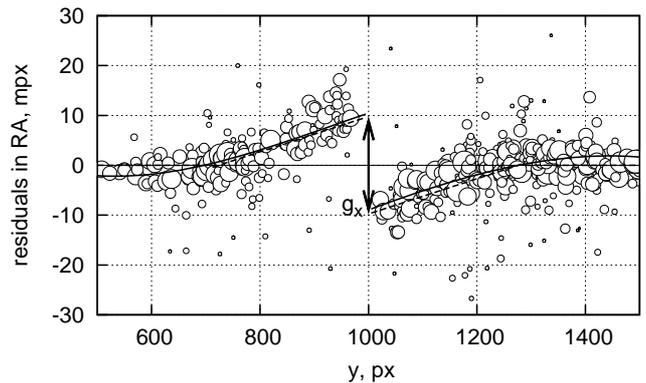}}
\caption{Individual epoch positional residuals $\Delta^x_{e}$ of stars in the calibration files  (open circles with sizes increasing with stellar brightness)  for a single epoch in the field of dw14,  its approximation Eq.\,(\ref{eq:G}) (solid lines) with $g_{x,e}=19.5$~mpx, and the astrometric model response $W_{CAL}(y)$ to the step signal $gU(y)$ in the RA positions of stars in  chip 2 (dashed lines). }
\label{dw14_ep5}
\end{figure}

 Fig.\,\ref{dw14_ep5}  shows the typical dependence  of the positional residuals   $\Delta^x_{i,e}$   on $y$ at the example epoch on 17 February 2012 for dw14. It has a discontinuity at $y=1000$~px and is approximated by the empirical function
\begin{equation}
\begin{array}{ll}
\label{eq:G}
G(y)  =\pm 0.5 g(1-\tilde{y}_i),    & \mathrm{if}  \hspace{2mm}     \tilde{y}_i<1 \\
G(y)  =0,                                         & \mathrm{if}   \hspace{2mm}    \tilde{y}_i>1  \\
\end{array}
,\end{equation}
where $\tilde{y}_i= |y_i-1000|/(0.35\,R_{\rm opt})$ is the normalized  distance from the chip gap, and a positive (negative) sign is used for stars in chip 2 (chip 1).  The fit function (\ref{eq:G})  shown in Fig.\,\ref{dw14_ep5} by solid lines  nearly coincides with the response function $W_{CAL}(y)$  (dashed)  reproduced from Fig.\,\ref{responce} and normalized to the offset amplitude $g=19.5 \pm 0.6$~mpx in the measured positional residuals. The estimates of $g_{x,e}$ in $x$ and $g_{y,e}$ in $y$  were derived  for each epoch $e=1 \ldots N_e$ by a least-squares fit of function  (\ref{eq:G}) to the residuals  $\Delta^x_{i,e}$, $\Delta^y_{i,e}$. The average uncertainty in the determination of the offset $g$  is $\sigma_g=1.3$~mpx.

\begin{figure}[h!]
\begin{tabular}{@{}c@{}}
\resizebox{\hsize}{!}{\includegraphics* [width=\linewidth]{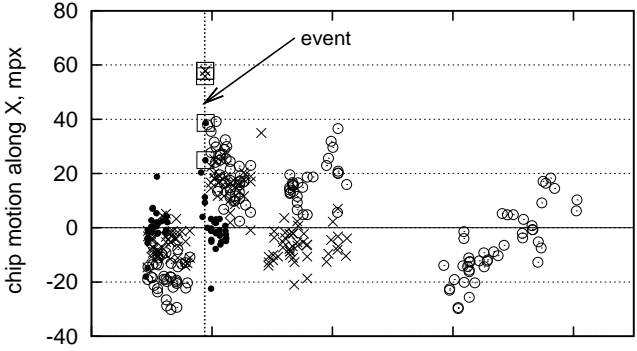}}\\
\end{tabular}
\caption{Change in time  of chip offset $g_{x,e}$  for sky fields with  short  (filled dots, six sky fields), medium  (crosses, seven fields), and long  (open circles, seven fields) observation time-span. The vertical line marks the time of an event that caused a 20--50~mpx offset in $g_{x,e}$ for four fields observed on 20--24 November 2011 (squares). }
\label{eppx}
\end{figure}

The change in $g_{x,e}$ values with time for all sky fields is  shown in Fig.\,\ref{eppx}.  The behavior is different  for  short- (one to one and a half years), medium- (three years), and  long-duration (seven years) datasets, which is expected:  the astrometric reduction uses the model Eq.\,(\ref{eq:eq1}), which fits the measurements of every star $i$ by the linear change in time caused by proper motion, hence the residuals $\Delta^x_{i,e}$ and $\Delta^y_{i,e}$ do not change linearly in time. The set of values $g$ for each star field  derived later by the least-squares fit Eq.\,(\ref{eq:G}) of these residuals are  linear combinations of these  residuals.  Therefore, a set of $g_e$ values for a given star field with its observation time-span also has zero linear dependence on time  within the corresponding time intervals,  as  seen in Fig.\,\ref{eppx}. In particular, this explains why the $g$ values for short-duration datasets show no trend in 2010--2012, while that is not the case for the longer-duration datasets.

The vertical line in Fig.\,\ref{eppx} marks the time of a large fluctuation on 20--24 November 2011 in four fields. This event is even more prominent in the reduced residuals (Fig.\,\ref{oc}).

\begin{figure}[h!]
\resizebox{\hsize}{!}{\includegraphics*[width=\linewidth]{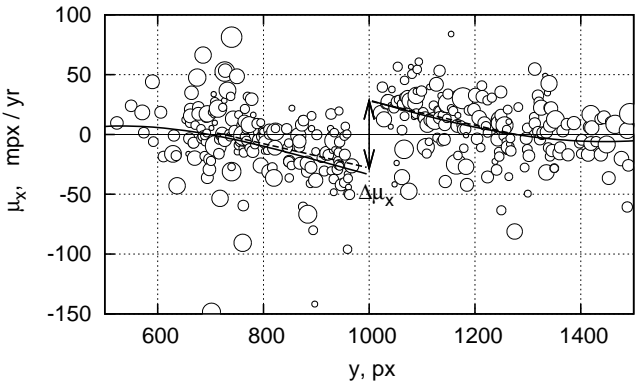}}
\caption{Same as Fig.\,\ref{dw14_ep5}, but  for proper motions:  the approximation of Eq.\,(\ref{eq:G}) (solid lines)  and the  model response $W_{CAL}(y)$ to the step change $\Delta \mu =-56.0$~mpx yr$^{-1}$ in the proper motions  of stars in chip 2 (dashed lines).}
\label{mu}
\end{figure}

The discontinuity of $W_{CAL}(y)$ type is seen in all $\xi$ parameters, for instance, in the parallaxes \citepalias[cf.][Fig.6]{PALTA2}.  Fig.\,\ref{mu} shows the distribution of the CAL dataset  proper motions for the example case of dw14, which is well approximated by the $W_{CAL}(y)$ function with an offset amplitude of $\Delta \mu =-56.0 \pm 5.0$~mpx yr$^{-1}$. The offset terms $g$ and $\Delta \mu$ can be caused by the RCM, but the nonzero offset values for the other parameters $\xi$ are unphysical; they represent the  formal fit solution of a model with correlated parameters.

\section{Chip motion retrieved from the calibration files}{\label{mot2}}
The definition in Eq.\,(\ref{eq:def_g}) means  that the parameter $g$ incorporates information on the epoch residuals of all  stars near the chip gap. It is therefore convenient to represent the full star sample for a particular sky field $z$  by one single 'equivalent' star that is located right at the chip gap on chip 2 and that has the epoch residuals  $g_{x,z,e}$, $g_{y,z,e}$. For this star, the equivalent  proper motions $\Delta \mu_{x,z}$, $\Delta \mu_{y,z}$ were derived by fitting  individual stellar motions $ \mu_{x,z,i}$, $ \mu_{y,z,i}$ with the function $W_{CAL}(y)$. The equivalent star displacement at epoch $e$ relative to time $t=0$ is 
\begin{equation}
\label{eq:V}
\begin{array}{ll}
V^x_{z,e} = g_{x,z,e} +\Delta \mu_{x,z} t_e, & V^y_{z,i,e} = g_{y,z,e} +\Delta \mu_{y,z} t_e.  
\end{array}
\end{equation}
The values of $V$ can be considered as the measurements in the right side of  the model   Eq.\,(\ref{eq:eq3}),  therefore  we rewrite this model into a system of $N_e \times N_{\rm dw} \times  2$ equations, 
\begin{equation}
\label{eq:eq}
\begin{array}{l}
 \sum_s \hat{\xi}_{z,s} \nu^x_{s,e} + H^x(t_e)  +  \delta H^x_{z,e}=  V^x_{z,e}, \\   
\sum_s \hat{\xi}_{z,s} \nu^y_{s,e} +  H^y(t_e)  +  \delta H^y_{z,e} = V^y_{z,e}, \\
\end{array}
\end{equation}
where $z=1 \ldots N_{\rm dw}$ is  the sky field index, $\hat{\xi}_{z,s}$ are the  formal fit parameters (offsets $\hat{x}_{z}$, $\hat{y}_{z}$, proper motion $\hat{\mu}_{x,z}$, $\hat{\mu}_{y,z}$, etc.), $H(t_e)$ is the smooth component of the RCM, and $\delta H_{z,e}$ are  the residuals of the fit, which include the measurement errors and the irregular component of the RCM. The fit parameters $\hat{\xi}_{z,s}$ are the corrections to the $\xi$ values in Eq.\,(\ref{eq:eq1}) derived from the primary astrometric solution with $H(t)=0$. The system of equations (\ref{eq:eq}) allows us to extract $H(t)$. When applied to a single sky field,  Eq.\,(\ref{eq:eq}) returns the trivial solution  $\hat{\mu}_{z}= \Delta \mu_{z}$, $\delta H_{z,e}= g_{z,e}$, and $H(t_e)=0 $. A meaningful solution is found by applying it to all fields. For the function $H(t)$ we adopted  the  low-order polynomial approximation 
\begin{equation}
\label{eq:H}
H(t) =  \sum_{j=0}^{n}  h_j  t^j
\end{equation}
with $n=3$ and  $n=1$ for the measurements along $X$ and $Y$, respectively. 
To separate the fully correlated proper motion and $h_1$ terms,  we  required that the sum of proper motions taken over all stars and  fields is  zero, that is,\ $\sum_z \hat{\mu}_{x,z}= \sum_z \hat{\mu}_{y,z} =0$. With a similar restriction on the stellar position offsets $\hat{x}_{z}$ and $\hat{y}_{z}$, we separated these terms and $h_0$. Thus, the systematic position change in time is fully included in the function $H(t),$ whereas the parameters $\hat{x}_{z}$, $\hat{y}_{z}$,  $\hat{\mu}_{x,z}$, and $\hat{\mu}_{y,z}$ have an average value of zero.

\begin{figure}[h!]
\resizebox{\hsize}{!}{\includegraphics*[width=\linewidth]{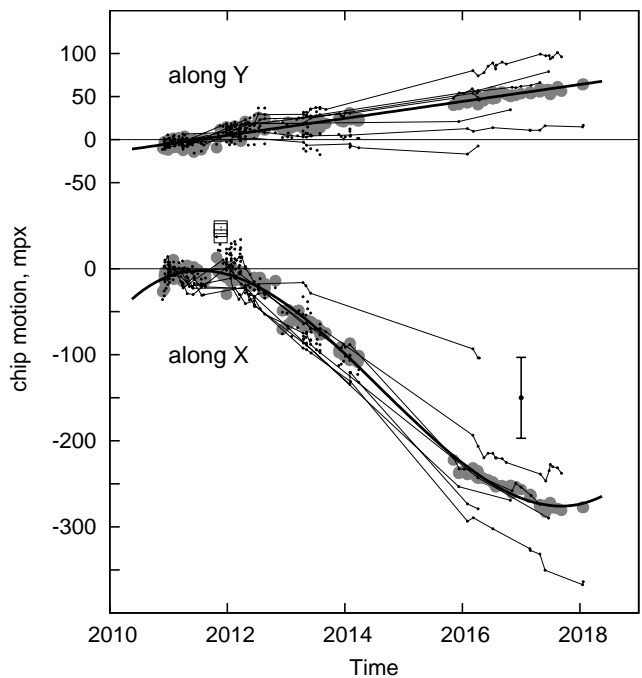}}
\caption{Motion of chip 2 along the $X$ -axis (lower panel) relative to a reference frame defined by stars in both chips: the unreduced displacements $V^x_{z,e}$ for each epoch $e$ and each sky field $z$ are shown by dots with the typical uncertainty in 2017 (error bar).  Solid lines connect the long-duration datasets. The model function $H(t)$ for the motion of chip 2 (thick solid line) and the instant chip motion $ H^x(t_e)  +  \delta H^x_{z,e}$  derived with Eq.\,(\ref{eq:eq}) (gray circles) are shown as well.  The large offset in four fields during  20--24 November 2011 is marked by squares.  The upper panel shows the same for the measurements along the $Y$ -axis.}
\label{abs_move2}
\end{figure}

Figure \ref{abs_move2}  shows the  unreduced displacements $g_{x,z,e}+ \Delta \mu_{x,z} t_e$, which reveals significant changes of the position of chip 2 relative to chip 1, which reaches  offset of 0.2--0.3 pixels in 2016--2018. These displacements are measured consistently in several fields and can therefore not be caused by particular stellar motions. The solution  $H(t)$  for the motion along the $X$ -axis is derived from Eq.\,(\ref{eq:eq}) as  a cubic function of time  (solid line in Fig.\,\ref{abs_move2}). It has an approximately linear segment between 2012 and 2017 where the rate is $\sim 50$~mpx yr$^{-1}$.  Along the $Y$ -axis, the motion is approximately linear at a rate of $\sim$10~mpx yr$^{-1}$.  

\begin{figure}[tbh]
\resizebox{\hsize}{!}{\includegraphics*[width=\linewidth]{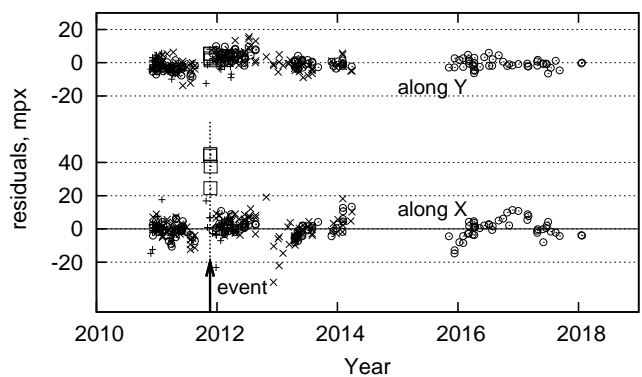}}
\caption{Fit residuals $ \delta H$ between the measured and the position of chip 2 according to the model  Eq.\,(\ref{eq:eq}) relative to chip 1 for each epoch and the sky fields with a short (filled dots), medium (crosses), and long (open circles) duration of observations. An impulse jump  for four fields obtained on 20--24 November 2011 is marked by squares.  }
\label{oc}
\end{figure}

Figure \ref{oc}  presents  the change in time of the fit residuals $\delta H_{z,e}$  for short-, medium-, and long-duration datasets. The residual distribution is approximately random with an RMS  of 4.3 and 3.8~mpx along the $X$ - and $Y$ -axes, respectively. Weak correlations and departures from the cubic model Eq.\,(\ref{eq:H}) can be seen, for example,\ close to 2017.0.  In general,  we did not find any strong  deviations on timescales of 1--4 days, which means that the RCM is quite smooth in time. 
 
The only strong variation on timescales of days is found for four fields observed on 20, 22, and 24 November 2011. The large 20--50~mpx offsets occurred within one day because two fields had been observed on November 19 with normal results. This event was probably caused by an intervention on the instrument as part of regular maintenance. Upon our enquiry, the ESO User Support Department confirmed that the FORS2 CCDs were switched on 28 November 2011 and that the observatory records do not indicate an earthquake recovery procedure or FORS2 cryostat warm-up or cool-down cycle on 20-24 November 2011, but that the exact dates of such interventions can sometimes be uncertain.

\section{Alternative measurements of the chip motion}{\label{mot1}}
{ In the past section we  retrieved the best precise displacement history of RCM using extensive data collected in the calibration files (the CAL dataset derived from multiple reference frames covering all FoVs), and with a reference frame $\omega$ set by stars in both chips. In order to confirm this primary result, we investigated the RCM by making several modifications to our astrometric solution. Instead of CAL files,  we now used the files  of RF datasets (stars of a single reference frame $\omega$ per each  sky field) and additionally restricted  $\omega$ space to chip 1. The motion of chip 2 stars then evidently reproduces the RCM effect.  However, we found that a robust  astrometric solution requires use of the normalizing parameter space $\omega^*$ expanded on both chips, in a field within $R_{\rm opt}$ distance from the dwarf. We also had to update the system of the normalizing conditions Eq.\,(\ref{eq:eq2}) by a model Eq.\,(\ref{eq:eqnew}) with  a flag $\lambda$ that allows for two options of the normalization (for details, see Sect.\,\ref{ch_7sub1}). 

Our computations confirm that the RCM does not depend significantly on the choice of the reference frame. Table\,\ref{table} summarizes the results of determining the RCM with these alternative approaches. }

\begin{table}[tbh]
\caption [] {Summary of the spaces  $\omega$,  $\omega^*$,  and flags  $\lambda$ that  specifies different modifications of astrometric reduction, and the estimated displacement of chip 2 along the $X$ -axis  from 2012.0 to 2017.5}
\centering
\begin{tabular}{@{}cccccccc@{}}
\hline
\hline
Nr&    file   &  $\omega$,  &  $\omega^*$,  & $\lambda$ & min $H^x$, &Sect.&Fig. \rule{0pt}{11pt}\\
    &   source &  chips               &    chips               &                         &        mpx                &   \\
\hline
1 & CAL       & 1\&2   & 1\&2       & 0  & -275 &  \ref{mot2}   &\ref{abs_move2},\,\ref{oc},\,\ref{fin}   \rule{0pt}{11pt} \\
2a & RF       &  1          & 1\& 2       & 1  & -308 &  \ref{ch_ee}     & \ref{abs_move},\,\ref{fin}\\
2 & RF          &  1         &  1\& 2      & 1  & -308 &  \ref{ch_dmu} & \ref{fin}\\
3 & RF          &  1         & 1\& 2      &  0  & -243 &  \ref{ch_dmu}  & \ref{fin} \\
4 & RF+DR2 & 1          & 1\&2     & 1  & -318 &  \ref{gaia}      &  \ref{fin},\,\ref{gaia_fig}\\
\hline
\end{tabular}
\tablefoot{Nr.2 a corresponds to using 15 bright stars in chip 2.}
\label{table}
\end{table}

\subsection{Motion of chip 2 derived in the reference frame of chip 1}{\label{ch_7low}}
To measure the displacements of chip 2, we performed the astrometric reduction Eq.\,(\ref{eq:eq1}--\ref{eq:eq3}) with the reference space $\omega$ limited to chip 1 by excluding its segment $\omega_2$  located on chip 2. We first also excluded $\omega_2$ from $\omega^*$ (Sect. \ref{ch_7sub1}), and we { revised }
 this option in Sect. \ref{ch_ee}.

\subsubsection{Systematic difference of proper motions between chips}
\begin{figure}[h!]
\resizebox{\hsize}{!}{\includegraphics*[width=\linewidth]{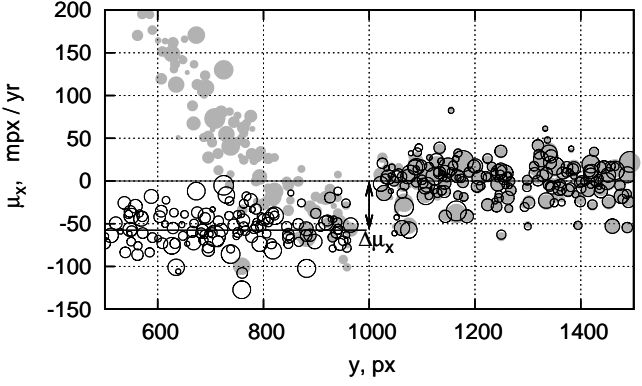}}
\caption{Proper motions  $\mu_x$  in the dw14 field derived relative to chip 1 (gray filled circles) and when the astrometric parameters are derived with Eq.\,(\ref{eq:eqnew}) (open circles), which allows for a step change between the chips  of $\Delta \mu_x=-56.0$~mpx yr$^{-1}$. The circle size is proportional to the stellar brightness.}
\label{step_mu}
\end{figure}

{\label{ch_7sub1}}
The dependence of  the solution coefficients $\xi$ on $y$, which we illustrate in Fig.\,\ref{step_mu} for proper motions,  is very different from the response function $W(y)$ shown in Fig.\,\ref{mu}.  The amplitude  of the discontinuity  $\Delta \mu_x$, however, is close to that derived with the calibration files.

For stars in chip 2 ($y<1000$ px), Fig.\,\ref{step_mu} shows a clear trend of $\mu_x$, which is natural because the restrictions Eq.\,(\ref{eq:eq3})  do not apply to stars in chip 2 in this case.  A flat dependence of $\xi$ on chip 2 is obtained by expanding  $\omega^*$ over this chip (i.e.,\ by including $\omega_2$ in  $\omega^*$) and allowing for a $\Delta\xi_s$ step change in the model  Eq.\,(\ref{eq:eq2}):
\begin{equation}
\begin{array}{l}
\label{eq:eqnew}
 \sum_{i \in \omega^*}  [\xi_{i,s}- \lambda  U(y_i) \Delta\xi_s] \bar {P}_{i} =0, \quad  s=1 \ldots S, \\
 \sum_{i \in \omega^*}  \xi_{i,s} \bar {P}_{i}  f_{i,w}(x_i,y_i) =0, \quad  s=1 \ldots S, \quad  w=2  \ldots  N'.
\end{array}
\end{equation}
Here 
\begin{equation}
\label{eq:rst}
\Delta\xi_s =\sum_{i \in \omega_2}  \xi_{i,s} \bar {P}_{i}  /  \sum_{i \in \omega_2}  \bar {P}_{i}  
\end{equation}
is a new  parameter that corresponds to the step change of $\xi$ in  chip 2. The $\lambda \in [0,1]$ parameter allows us to switch this addition on and off. This new model includes the offsets ($\Delta_x$, $\Delta_y$, $\Delta \mu_x$, $\Delta \mu_y$,  $\Delta \rho$,  $\Delta d$,  $\Delta \varpi$) of the parameter values between chips and excludes their systematic change across the FoV.

 Figure \ref{step_mu} shows that this approach works because the distribution of $\mu_x$ along $Y$ is flat, except for the step change of $\Delta \mu_x=-56.0 \pm 5.0$~mpx yr$^{-1}$ , which is very similar to that obtained from the calibration files, and because the proper motions computed with  Eq.\,(\ref{eq:eqnew}) (open circles) and  when the $\omega^*$ space is restricted to chip 1 (gray circles) are nearly identical for $y>1000$~px.

\subsubsection{Motion of individual stars in the lower chip}
\begin{figure}[tbh]
\resizebox{\hsize}{!}{\includegraphics*[width=\linewidth]{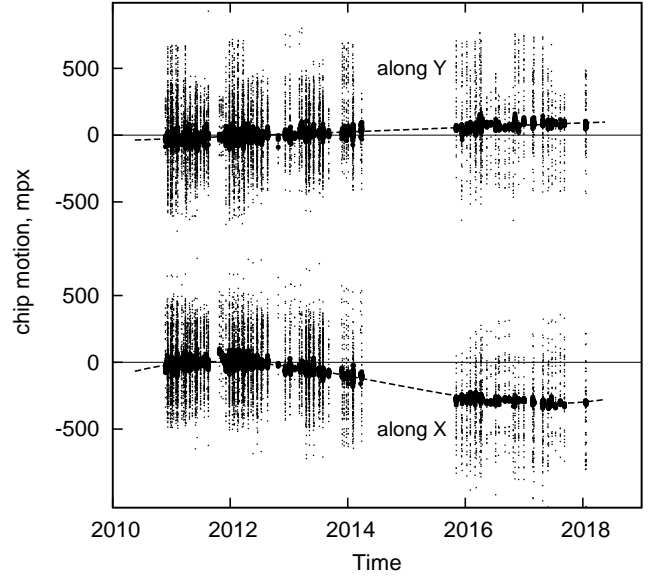}}
\caption{Motion of chip 2 relative to the reference frame of the upper chip:  unreduced  displacements $V_{z,i,e}$ along the $X$ - and $Y$ -axes of 15 bright lower chip stars in each sky field at each epoch (small dots). The same after correction for the residual parameters $\hat{\xi}$ (filled dots). { The model  motion} $H(t)$  (dashed line), which at this scale looks the same  for both the unreduced and the corrected displacements.}
\label{abs_move}
\end{figure}

{\label{ch_ee}}
With a reference frame anchored in chip 1, the space $\omega^*$ expanded to both chips, and the restrictions of Eq.\,(\ref{eq:eqnew}) with  $\lambda=1,$ we obtained a well-behaved step change in the parameters $\xi$ (a typical example is shown in  Fig.\,\ref{step_mu} with open circles). With this model, we ran a primary reduction with Eq.\,(\ref{eq:eq1}) and computed  the displacements 
\begin{equation}
\label{eq:VV}
\begin{array}{ll}
V^x_{z,i,e} = \Delta^x_{z,i,e} + \mu_{x,z,i} t_e, & V^y_{z,i,e} = \Delta^y_{z,i,e} + \mu_{y,z,i} t_e  
\end{array}
\end{equation}
 from the initial position of each star. Here $\Delta_{z,i,e}$ are the positional residuals, and $\mu_{z,i}$ are the proper motions of the stars $i$ in the field $z$. The displacements $V_{z,i,e}$ do not show any systematic trending in time for stars in the upper chip,  but reveal a clear trend in the lower chip, especially in $V^x_{z,i,e}$. Figure \ref{abs_move} shows the displacements $V^x_{z,i,e}$ for 15 bright but not saturated stars  in each field within a narrow $750<y<1000$~px area close to the chip edge. We defined homogeneous  star samples and discarded stars with proper motions $\mu_x$, $ \mu_y$ that deviated strongly from the average $\Delta \mu_x$, $\Delta \mu_y$ values  in a given field. 

The sample of data points $V_{z,i,e}$ presented here directly describes the motion of stars in chip 2 relative to the reference frame anchored in chip 1. Because the displacements are synchronous in all fields, we attribute this effect to the real RCM  rather than to peculiar motions of stars. We directly fit the full sample of individual stellar displacements $V^x_{z,i,e}$ along the  $X$ -axis  with a cubic function of time  and obtained a function  (dashed line in Fig.\,\ref{abs_move}) that is similar in  shape to the previous estimate of $H(t)$  in Fig.\,\ref{abs_move2}.

The values of $V_{z,i,e}$ vary as a result of the intrinsic proper motions of individual stars, and therefore Fig.\,\ref{abs_move} appears noisy. To obtain clearer results, we had to mitigate the impact of proper motions. For this purpose we applied Eq.\,(\ref{eq:eq}) as a secondary reduction model and replaced the right-hand side of the equations by  $V^x_{z,i,e}$ and $V^y_{z,i,e}$ from Eq.\,(\ref{eq:VV}).  This created a system of  $15 \times N_e \times N_{\rm dw} \times 2 $ equations with  polynomial order $n=3$  in $H^x(t)$ and  $n=1$  in   $H^y(t)$. We solved  it under the $s=1 \ldots 7$ restrictions of $\sum_{z,i} \xi_{z,i,s}=0 $. The obtained epoch displacements $ H^x(t_e) +\delta H^x_{z,i,e}=V^x_{z,i,e}  - \sum_s \hat{\xi}_{z,i,s} \nu^x_{s,e}$  shown by the solid dots in Fig.\,\ref{abs_move}  differ from the raw displacements $V^x_{z,i,e}$ only by the single term $\sum_s \hat{\xi}_{z,i,s} \nu_{s,e}$. These new estimates are tightly correlated with the direct approximation of $V^x_{z,i,e}$, $V^y_{z,i,e}$ obtained from 15 bright stars (dashed lines in Fig.\,\ref{abs_move}).  The scatter decreased from the initial $\pm 200$~mpx in  $V_{z,e}$ values  to $\sim 20$~mpx in the residuals $\delta H_{i,e}$, mainly because we subtracted individual stellar proper motions from the residuals. We found  that  the RCM is common for all sky fields, as expected, and that deviations from the smooth change in time are smaller than $\sim 20$~mpx.

\subsubsection{Chip  motion based on the properties of the stellar group}{\label{ch_dmu}}
To improve upon the results when 15 stars were used in the previous Section, we formed an equivalent star for each sky field $z$ in the way  described  in Sect.\,\ref{mot2}. This artificial star was positioned at the edge of chip 2, represents an ensemble of stars, and is characterized by the displacements
\begin{equation}
\label{eq:VVV}
\begin{array}{ll}
V^x_{z,e} = \Delta^x_{z,e} + \Delta \mu_{x,z} t_e, & V^y_{z,e} = \Delta^y_{z,e} +\Delta \mu_{y,z} t_e  
\end{array}
\end{equation}
measured relative to the reference frame of chip 1.  Here, $\Delta_{z,e}$ are the positional residuals, and $\Delta\mu_{z}$ are the proper motions of the equivalent star in a field $z$. In Sect.\,\ref{mot2}, similar estimates were derived from the fit of individual stellar parameters and their variation along $Y$ with the functions $W_{CAL}(y)$ and $G(y)$. We again used a step function (Fig.\,\ref{step_mu}) fit of individual stars corresponding to the parameter change along $Y$, derived the estimates at $y=1000$~px separately for each chip, and finally obtained  $\Delta_{z,e}$ and $\Delta \mu_{z}$  as the difference between the results in chip 2 and chip 1. 

\begin{figure}[tbh]
\resizebox{\hsize}{!}{\includegraphics*[width=\linewidth]{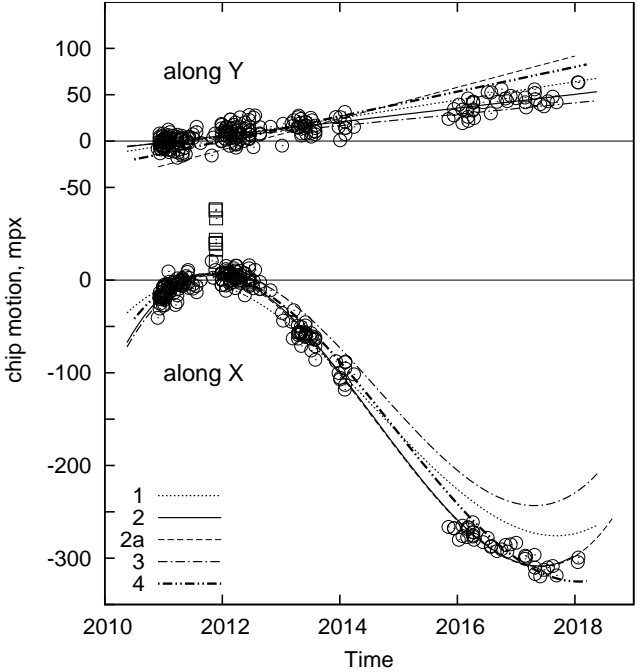}}
\caption{Motion of chip 2 along RA (lower panel) and Dec (upper panel) relative to chip 1.  Each curve is the solution for $H(t)$ derived with the models in Table\,\ref{table}, as indicated by the legend. The instant  motion estimate $H(t)+\delta H_{z,e}$ for  model 2 (open circles) shows the scatter in the RCM determination  for individual fields and epochs.  The step change in four fields observed on 20--24 November 2011 and derived with approaches 2 and 4 is marked by squares.  }
\label{fin}
\end{figure}

Then, using the  model Eq.\,(\ref{eq:eq}), we derived the  residual fit parameters $\hat{\xi}_{z,s}$, the functions $H^x(t)$,  $H^y(t)$, and the fit residuals $ \delta H^x_{z,e}$, $ \delta H^y_{z,e}$, whose rms  of 10~mpx is twice smaller than when individual stars are used (Sect.\,\ref{ch_ee}). The instant displacements $H(t)+ \delta H_{z,e}$  for each sky field and epoch  are  shown in Fig.\,\ref{fin} with open circles. The function $H^x(t)$ (solid curve)  is close to those obtained in the previous sections with 15 bright stars per field, and to $H^x(t)$ derived using both chips as reference field (Fig.\,\ref{abs_move2}). The differences between the estimates of $H^y(t)$ are small considering the uncertainty in proper motions. For both reduction versions the displacement amplitude of chip 2 is $\sim 0.25$~px.

We obtained one more estimate of $H(t)$  with the reference frame anchored in chip 1, where we used the full FoV as a space $\omega^*$ for the normalization  Eq.\,(\ref{eq:eqnew}--\ref{eq:rst}) of parameters ${\xi}$, but instead of $\lambda=1,$ we set  $\lambda=0$ (Nr.3 in Table\,\ref{table}). Now, the  parameters ${\xi}$ follow a  change that is approximated by the functions $W_{CAL}(y)$ or $G(y)$. Using these functions to fit the individual stellar parameters, we formed the equivalent stars for each sky field described by the group parameters $\Delta_{z,e}$ and $ \Delta\mu_{z}$. A subsequent  reduction  derived the model motion $H(t)$ that is shown by the dash-dotted curve in Fig.\,\ref{abs_move2} and   $\sim 10$~mpx  for the rms of fit residuals $ \delta H_{z,e}$.

\subsection{Using {\it Gaia} DR2 to estimate the chip motion}{\label{gaia}}
We used the {\it Gaia} DR2 catalog \citep{DR2, 2018yCat}  as an absolute calibrator to convert differential FORS2 astrometry into the ICRF coordinate system (ICRS). In this system, the RCM is apparent as a change in the astrometric zero-points of the CCD chips. To measure this effect, we converted FORS2 into ICRS  at every observation epoch. For the seven fields with the longest time-span, we have 147 epoch measurements, thus we obtained 147 estimates of the relative chip position in 2010--2018. The CCD epoch positions 
\begin{equation}
\label{eq:xep}
\begin{array}{lc}
\bar{x}_{z,i,e} =x_{z,i,0} + V^y_{z,i,e},  & \bar{y}_{z,i,e} =y_{z,i,0} + V^y_{z,i,e} 
\end{array}
\end{equation}
converted into ICRS are formed from the $i$-th star position $x_{z,i,0}$, $y_{z,i,0}$ in the reference image ($m=0$) and the displacements  $ V^x_{z,i,e}$, $ V^y_{z,i,e}$ computed with Eq.\,(\ref{eq:VV}) for each sky field $z$. Thus, we used  $\Delta_{z,e}$  and $\Delta \mu_{z}$ as previously derived with the astrometric reduction based on the reference frame $\omega$ limited to  chip 1, and the space $\omega^*$ set to all FoVs (Nr.\,4 in Table\,\ref{table}). 

The zero-point estimates potentially depend on the exact sample of stars used at a given epoch. We mitigated this by using the stars that are common to all epochs. This common star list was derived  at the first identification between FORS2 and {\it Gaia} at the epoch  nearest to the {\it Gaia} reference epoch $T_{DR2}=2015.5$.    The identification  procedure was similar to that described in \citet{gaia,2018LUH}, where we used  full bivariate  polynomials of the maximum power  $n$ for the transformation between catalogs. We used stars in both chips for this transformation. To take the relative displacement of  chip 2 into account, we expanded these transformation functions with two additional free parameters $\Delta F^x_{e,z}$ and $\Delta F^y_{e,z}$  that we added to the zero-points of chip 2 only and set to zero for chip 1. This model integrates information on the relative position of the reference frames and thus on the relative motion of chips. Upon the estimation of  $\Delta F_{e,z}$, we find the instant displacements
\begin{equation}
\label{eq:15}
H^x_z(t_e)+ \delta H^x_{z,e} = \Delta F^x_{e,z} - \overline{\Delta F}^x_{z} \quad H^y_z(t_e)+ \delta H^y_{z,e} = \Delta F^y_{e,z} - \overline{\Delta F}^y_{z}, 
\end{equation}
where $\overline{\Delta F}_{z} $ is the average offset in a field $z$ on 2011--2012.
We tested all polynomial orders $n$ up to  the maximum $n=7,$ and its best-fit value was found using an F-test for the rejection of insignificant coefficients as described in \citet{gaia}. We found that  $n=5$ fit most fields best.

We set the cross-match radius to $3.3\sigma_1$, where $\sigma_1$ is the model uncertainty that includes all known errors of FORS2 and {\it Gaia} DR2, and is 1--2~mas for G=16--19~mag stars at the  epochs near $T_{DR2}$. In different fields we identified between 43 to 206 stars that FORS2 and DR2 had in common.  For the  epochs $\pm 0.5$~yr within $T_{DR2}=2015.5$,  we obtained the unbiased rms of individual differences between FORS2 and {\it Gaia} positions  $\sigma_{F-G}=1-2.5$~mas, which is close to the expected value of $\sigma_1$.   The dominant error component of $\sigma_1$ is the uncertainty of {\it Gaia} proper motions of about  { 0.3--1~mas~yr$^{-1}$}. For this reason,   $\sigma_{F-G}$ is minimum on $T_{DR2}$  and  increases   to 4--7~mas at the  2011 and 2017 epochs.

\begin{figure}[tbh]
\resizebox{\hsize}{!}{\includegraphics*[width=\linewidth]{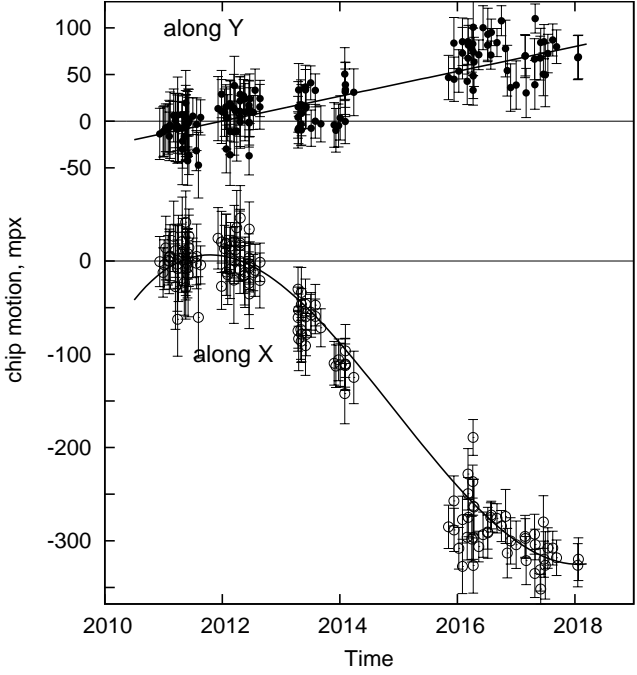}}
\caption{Motion of chip 2 along RA (lower panel) and Dec (upper panel) relative to  the upper chip derived using {\it Gaia} DR2:  Instant estimates $H_z(t_e)+ \delta H_{z,e}$ for individual sky fields $z$ and epochs $e$  (open circles) and the smooth fit functions H(t).}
\label{gaia_fig}
\end{figure}

In some fields we found fewer than expected cross-matched stars in chip 2, and the number was too low to determine $\Delta F_{e,z}$ reliably.  Better results were obtained when the common star list was created from the identification made at the first epoch of the sky field exposure (2011--2012). For these epochs, which are  distant from $T_{DR2}$, a  higher  value  of $\sigma_1$ and thus a wider  identification window   favorably   increased  the common  star  number. With this approach, the $\sigma_{F-G}$ value increased slightly  but the uncertainty in the determination of $\Delta F_{e,z}$   improved   significantly in some fields. Finally, we obtained the values $\Delta F_{e,z}$ and computed  the instant motion of chip 2 $H_z(t_e)$ that we present in Fig.\,\ref{gaia_fig} with Eq.\,(\ref{eq:15}). These estimates can be approximated by the polynomials
\begin{equation}
\label{eq:16}
\begin{array}{l}
H^x(t)= -204.5 - 76.78 t +4.66 t^2 +2.701 t^3  \hspace{2mm}  \mathrm{mpx},\\
H^y(t)= 46.7 +13.31 t   \hspace{2mm} \mathrm{mpx} \\
\end{array}
\end{equation}
with uncertainties of $\pm 0.08$ and  $\pm 0.04$~mpx for the linear coefficients in RA and Dec, respectively, and $t$ is measured relative to $T_{DR2}$. In 2017.5,  $H^x=-318$~mpx and $H^y=67$~mpx, which matches the estimates derived with the other approaches (Table\,\ref{table}) relatively well.

\section{Calibration of and correction for the chip motion}{\label{alt}}
The relative displacement of the chips directly affects the  position of the target by adding a bias to the epoch residuals  $\Delta_{z,e}$   (see Fig.\,\ref{dw14_ep5}). According to Eq.\,(\ref{eq:G}), the bias value  $G(y)$  depends on the amplitude $g$,  the target distance from the  division line between chips, and  the reference field size $R_{\rm opt}$.  Typically, $g_x=10-20$~mpx and  $R_{\rm opt}= 500-1000$~px, which for a target located at $y=1100$~px results in an RA bias of $\approx $3--6~mpx or 0.4--0.7~mas at a single epoch. This is a rough  estimate for the rms variation of the bias added to the epoch residuals, which we denote as $\sigma_{\mathrm{bias}}$ and which is  very large in comparison to the  FORS2 astrometric uncertainty of $\sim$0.1~mas. In the extreme events of $g \approx 60$~mpx, the bias in these epoch residuals can reach  2~mas.  This bias therefore needs to be identified, modeled, and removed. 

The bias can be removed in two ways:  either using the discrete estimates $g_e$  derived from the calibration files, or using the analytic approximations for $H(t)$ derived in Sects.\,\ref{ch_7low} and \,\ref{gaia}. In the first case, incorrectly assuming that the RCM  is  a random process that is not correlated in time, we expect that $H(t_e)=g_e$, so that a simple correction   $x_m - g_{x,e}$,   $x_m- g_{y,e}$ applied to the measurements of stars in chip 2 eliminates the effect. 

\begin{figure}[tbh]
\resizebox{\hsize}{!}{\includegraphics*[width=\linewidth]{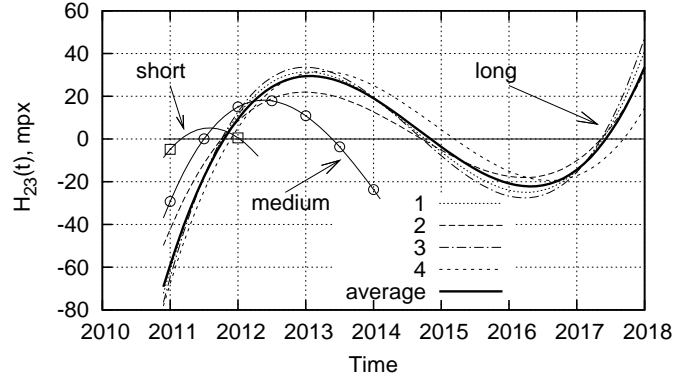}}
\caption{ Nonlinear component $H^x_{23}(t)$ of the  displacement of chip 2 in time derived with approaches 1--4 listed in Table \ref{table}, and their average $\langle H \rangle$ (thick solid line) for the long-duration datasets (${\Delta t}=7$~yr ), and $H^x_{23}(t)$ corresponding to  $\langle H \rangle$ for the short-duration (${\Delta t}=1.5$~yr) and medium-duration (${\Delta t}=3$~yr) datasets.}
\label{non_lin}
\end{figure}

However,  the values of $H(t_e)$ are clearly correlated and reflect the mostly smooth motion of the chips. The  models of $H(t)$ derived in Sects.\,\ref{mot2} and\,\ref{ch_7low} are sums of the linear  and  nonlinear components $H_{23}(t)$ with quadratic and cubic ($t^2$ and $t^3$) terms.  The linear component is included in the proper motions of stars and therefore does not affect the computation of the epoch residuals and $g$ values.  The following discussion therefore does not concern the RCM along $Y,$ which is linear in time. 

The component  $H_{23}(t)$ produces RCM variations of $\pm 20-50$~mpx amplitude and correlated on a 1--2~yr timescale (Fig.\,\ref{non_lin}). The astrometric reduction  redistributes the  $H_{23}(t)$ variations across the model parameters $\xi$ and the epoch residuals. Moreover, because some sky fields are observed within a restricted time interval $\Delta t$,   instead of $H_{23}(t),  $ we detected only the remnants $\hat{H}_{23}(t)=H(t)-a_{\Delta t}-b_{\Delta t} t$ of this function after subtracting the linear fit $a_{\Delta t} +b_{\Delta t} t$ over the time interval ${\Delta t}$ specific to a field. The shape of $H_{23}(t)$ therefore depends on $\Delta t$, and the variation amplitude increases with $\Delta t$. Thus, $H(t)$ is not fully reflected in the epoch residuals, and we consequently measure a reduced value of $g_e$. 

\begin{figure}[tbh]
\resizebox{\hsize}{!}{\includegraphics*[width=\linewidth]{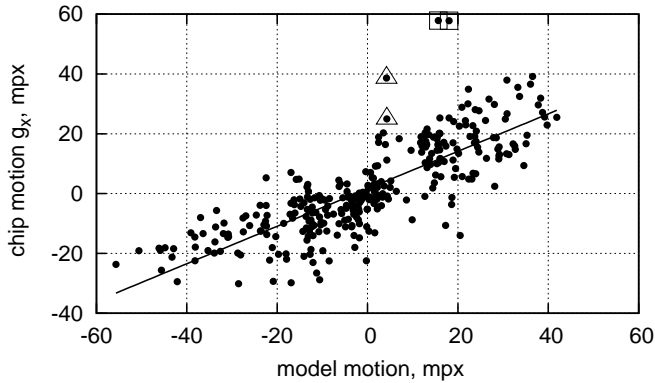}}
\caption{Correlation between the epoch displacements in chip 2,   $g_x,$ along RA  in 20 sky fields derived with the standard reduction (Nr.1 in Table\,\ref{table}) and the nonlinear component of the chip motion $\hat{H}_{23}(t)$ derived with use of {\it Gaia} DR2 (Nr.4 in Table\,\ref{table}). The fitted line has a slope of $K=0.63 \pm 0.03$.  The measurements on 20--24 November 2011 are shown by squares and triangles for the medium- and short-duration datasets, respectively.}
\label{gg}
\end{figure}

This is illustrated with Fig.\,\ref{gg}, where we compare the measured $g_{x,z,e}$ values derived from the CAL files for each sky field $z$ and epoch $e$, and  $\hat{H}_{23}(t)$  derived using {\it Gaia} DR2 and explicitly set  by  Eq.\,(\ref{eq:16}) for $H(t)$. This graph shows that the $g$ values are on average lower than the  expected $\hat{H}_{23}(t)$ values.  We find a linear  dependence
\begin{equation}
\label{eq:K}
g_e = K  \hat{H}_{23}(t)
\end{equation}
between these values  with a slope of $K=0.63 \pm 0.03$.  Thus, given the measured $g_e$ value, the statistical RCM is $\hat {H}_{23}(t_e)= (1/K) g_e=(1/0.63) g_e$, which is close to the empirical $(1/0.70) g_e$ estimate derived in \citetalias{PALTA2}. Similar estimates for $K$ were obtained with other models of RCM, for instance,\  $0.62 \pm 0.03$ and  $0.58 \pm 0.02$  for solutions 2 and 3 of Table\,\ref{table}. With the correction  
\begin{equation}
\label{eq:17}
x_m^{\mathrm{corr}} = x_m - K^{-1} g_{x,e},   \quad y_m^{\mathrm{corr}}=y_m-  K^{-1}g_{y,e}
\end{equation}
applied to the measurements of stars in chip 2,  the  systematic change in residuals with $Y$ similar to that shown in  Fig.\,\ref{dw14_ep5} is removed \citepalias[Sect.4.3.2, Fig.5]{PALTA2}.   Given the average uncertainty $\sigma_g =1.3$~mpx of the $g$ determination, the effect of RCM in the epoch residuals of  targets at $y_0=1100$~px  decreases by one order of magnitude from the initial  $\sigma_{\mathrm{bias}}=$3--6~mpx  to $\sigma_{\mathrm{bias}}=  K^{-1} \sigma_g  g^{-1}G(y_0) \approx 0.6$~mpx or $\pm$0.07~mas. This residual bias is acceptable for FORS2 astrometry.

Alternatively, the RCM  can be removed directly using one of the derived model functions $H(t)$. As explained above, it is sufficient to apply these corrections for the nonlinear motion components only because the linear component (i.e.,\ the case of Dec) does not affect the epoch residuals. In the previous sections we obtained a few estimates of $H(t)$ whose  nonlinear components $ H_{23}(t_e)$  (Fig.\,\ref{non_lin}) are slightly dispersed, and we formed their average   $\langle H \rangle$ to obtain more reliable results. The  rms  of  the individual $ H_{23}(t_e)$ relative to $\langle H \rangle$ is 6.1~mpx, therefore the  formal uncertainty of  $\langle H \rangle$ is $\sigma' \approx 3.1$~mpx. We can now correct for RCM by applying the correction
\begin{equation}
\label{eq:19}
x_m^{\mathrm{corr}} = x_m -  \langle H(t_e) \rangle
\end{equation}         
to stars in chip 2. For stars at $y_0=1100$~px, the bias is expected to decrease to  $ \sigma_{\mathrm{bias}}=\sigma' g^{-1}G(y_0) \approx \pm 0.12$~mas, which is larger than the residual bias obtained when $g_e$ is used.

We verified the above estimates with the complete astrometric reduction of dw14 by including and excluding the RCM corrections Eqs.\,\ref{eq:17} and \,\ref{eq:19}. We applied a reduction model based on  Eq.\,(\ref{eq:eq1}) and Eq.\,(\ref{eq:eq2}) (or, equivalently,  Eq.\,(\ref{eq:eqnew}) with $\lambda=0$) using  circular $\omega$ and $\omega^*$ spaces  of $R_{\rm opt}=711 $~px radius centered at dw14.   This is a standard model  for  astrometric reduction on FORS2 \citep{PALTA1, PALTA2}, but now we take into consideration that $\omega$ and $\omega^*$ spaces are fragmented into two sections that are related to different chips.   

\begin{figure}[tbh]
\resizebox{\hsize}{!}{\includegraphics*[width=\linewidth]{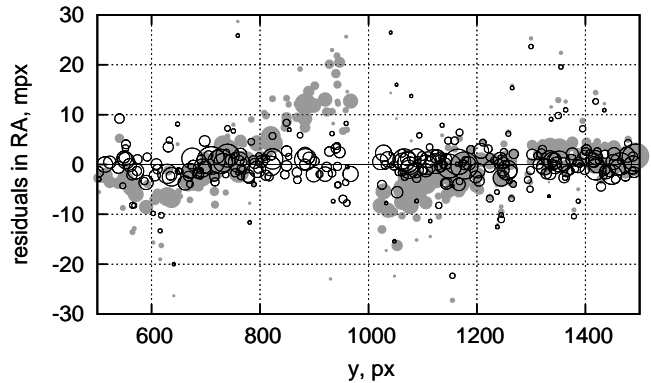}}
\caption{Epoch residuals  $\Delta_{e}^x$  for reference stars in the field of dw14 obtained in a single epoch (same as in Fig.\,\ref{dw14_ep5}) without correction for RCM (gray circles) and after correction with Eq.\,(\ref{eq:17}) and $K=0.63$ (open circles). }
\label{both}
\end{figure}

Without the RCM correction, the astrometric reduction produced the epoch residuals  $\Delta_{e}^x$  shown in Fig.\,\ref{both} by gray circles for the example epoch on 17 February 2012, which can be compared to  Fig.\,\ref{dw14_ep5} at the same epoch, but based on the calibration files (Sect.\,\ref{datasets}). The plots are similar in shape, but the distribution of  $\Delta_{e}^x$ in Fig.\,\ref{both}  is now not exactly symmetric 
relative to $y=1000$~px  because the reference field center is 73~px off this line. Moreover, the number of stars is twice smaller than in the calibration files because of the limited  $R_{\rm opt}$ size. 

The  systematic   value of $\Delta_{e}^x$  at some $y$ is the bias in the position of the object located there, so that for dw14  at  $y=1073$~px  the bias  in the example epoch  is roughly   $0.28g_{x,e} \approx$5~mpx or 0.6~mas. The rms value  $ \hat {\sigma}_{\mathrm{bias}}$ of the bias taken over all epochs is 0.55~mas. This includes the uncertainty of the fit $\sigma_{fit}= 0.06$~mas, which we subtracted quadratically and obtained $ \sigma_{\mathrm{bias}}=0.54$~mas. This value characterizes the bias in the epoch residuals  without correction of RCM (Table\,\ref{table2}).

The efficiency of the correction with Eq.\,(\ref{eq:17}) is evident from  Fig.\,\ref{both}, which displays the new residuals $\Delta_{e}^x$ (open circles) computed with the positions $x_m^{\mathrm{corr}}$,  $y_m^{\mathrm{corr}}$ of stars in chip 2 corrected for each image $m$. The distribution of  $\Delta_{e}^x$  along $Y$ is now flat.  For the example epoch and objects at $y=1073$~px, this calibration decreases the bias to 0.3~mpx.  For all epochs,  the average rms bias for the target is $ \hat{\sigma}_{\mathrm{bias}} =$0.06--0.07~mas; it is equally good for the two tested coefficient values of $K=0.70$ and  $K=0.63$. By quadratically subtracting  $\sigma_{fit}$ , we find $ \sigma_{\mathrm{bias}} =$0.02~mas for the correction with $K=0.63$. 

This is the best way of removing the RCM effect and reducing its impact to the noise  level. We also find  that  for the epoch shown in  Fig.\,\ref{both} this calibration decreased  the rms of $\Delta_{e}$ for bright 17--18~mag  stars from 4.2~mpx (=0.5~mas) to  1.0~mpx (=0.12~mas), which is a typical  uncertainty of the epoch residuals for FORS2 astrometry.  

A significant but insufficient decrease of  $ \sigma_{\mathrm{bias}}$ to 0.16~mas (Table\,\ref{table2}) was also obtained with the calibration of Eq.\,(\ref{eq:19}). This result indirectly suggests that the true RCM follows a smooth cubic polynomial change in time, but that the updated model of $H(t)$ should include higher order terms in $t$ and possibly some irregular displacements like those that took place on 20--24 November 2011.
For comparison, we repeated these computations for the ultracool dwarfs 3 and 11 listed in Table\,1 by \citet{ PALTA1}, which represent short- and  medium-duration datasets, respectively. 

\begin{table}[tbh]
\caption [] {Rms of additional bias  $ \sigma_{\mathrm{bias}}$ in the epoch residuals in mas obtained  with and without RCM correction, for the short-, medium-, and long-duration datasets indicated by $\Delta t$.}
\centering
\begin{tabular}{@{}c|ccc@{}}
\hline
\hline
 & \multicolumn{3}{|c}{$\Delta t$}\\
correction type                & short & medium    & long   \rule{0pt}{11pt}\\
\hline
no correction                 &       0.11  &0.53 & 0.54  \rule{0pt}{11pt} \\
 Eq.\,(\ref{eq:19})          &      0.08  &0.22 &0.16\\
 Eq.\,(\ref{eq:17})         &      0         & 0 &0.02\\
\hline
\end{tabular}
\tablefoot{Eq.\,(\ref{eq:17}) uses discrete estimates $g_e$ derived from calibration files for each epoch $e,$ while Eq.\,(\ref{eq:19}) uses functional expressions for the chip motion in time. }
\label{table2}
\end{table}

Summarizing these results, we conclude that the RCM effect is comparable for long- and medium-duration datasets and can be removed well using $g$ values based on calibration files. For short-duration datasets, the RCM adds a small bias that does not significantly degrade the astrometric precision even when left uncorrected.

\section{Conclusion}{\label{conc}}

Astrometric programs often require a reference field whose size exceeds a single chip. For multi-chip detectors it is therefore natural to use stars both in the primary chip where the target object is imaged and  in the adjacent chips. In the case of FORS2, the reference field partially covers the primary chip 1, but for precise astrometric reduction  it is often extended to the second chip 2. The use of such multi-chip reference frames implies that the locations of the chips within a detector are known.

First signs of internal chip motions within the FORS2 detector were reported in \citetalias{PALTA2}. Here, we investigated this effect in detail using images of sky fields that were monitored over a few years \citep{PALTA1}. We used several approaches: the motion of chip 2 was measured relative to chip 1 or relative to the reference frame extended over both chips, and we used the {\it Gaia} DR2 catalog. All approaches yielded concordant results in which the motion of chip 2 varied in time with an amplitude of up to $\sim 70$~mpx yr$^{-1}$. The full amplitude of the chip displacement between 2011 and 2018 is 0.3~px or 38~mas in the direction along the line that divides the chips. In the orthogonal direction (i.e.,\ across the chip gap) the displacements is $\sim 0.07$~px . 

The dominant relative chip motion follows a cubic dependence on time in RA and changes its direction in 2011.8 from positive to negative, and it seems to invert again in 2017.5.  In addition, we found significant deviations from a smooth continuous motion. For example, we detected a step-like displacement by 0.02--0.05~px (2--6~mas) that occurred on 20--24 Nov. 2011 and was likely caused by an instrument maintenance intervention.   In the orthogonal direction, the relative chip motion appears linear in time. 

On average, the relative chip motion adds a random component to the epoch residuals of $\sim$$0.5$~mas, thus it can significantly degrade the FORS2 astrometric performance of 0.1~mas for a single epoch if it is not corrected for. We demonstrated that the relative chip motion effect can be mitigated to the noise level at the cost of extensive computations that are needed to create special calibration files.

The relative chip motion was investigated by \citet{dark}, who analyzed the stability of a 61 CCD chip array of the Dark Energy Camera (DECam) \citep{DECam} and discovered displacements of chips with an amplitude of 0.4~px=100~mas, or 6~$\mu$m. Although these estimates are comparable to those we obtained for FORS2, they refer to  the detectors of a quite different design and linear dimensions.  The DECam CCDs cover a very large $\sim 450 \times 450$~mm  space, so that the   displacement values are $10^{-5}$ relative to the detector size only, while we found  $\sim 10^{-4}$ for FORS2. The  motions of the DECam CCDs were found to be closely related to the warming or cooling thermal recycling, which appears as a stochastic change in time. In contrast, the motion of FORS2 chips{ within the MIT detector} is smooth and linear in time, at least between 2012 and 2018, and does not depend much on the thermal recycling during the exchanges between the MIT and  E2V mosaic detectors, which occur { with a spacing of a few months.} 
For these reasons, we consider that the relative chip motion in the case of FORS2 is of a different physical nature and potentially reflects some aging processes.

We demonstrated that chip array instabilities can seriously degrade high-precision astrometry if left uncorrected. The MICADO near-infrared camera \citep{MICADO2010} for the ELT aims at a single-epoch  accuracy of $\leq 50$~$\mu$as  over  an FOV  of  53\arcsec  \citep{MNRAS_ELT, SPIE_ELT}. The detector of MICADO is a $3\times 3$ array, and  if the reference field is expanded over the full FoV, the RCM effect can  fragment it into  $3\times 3$ segments. In the hypothetical case that the relative chip motion of MICADO is comparable to FORS2, it can introduce biases of 50--300~$\mu$as in the target epoch positions if the reference field extends over several chips and the effect is not calibrated and corrected  using dedicated methods (e.g., \citet{Rod2019}).

Our findings underline that chip stability needs to be considered. Dedicated monitors for detector arrays may need to be implemented that are expected to fulfill the stringent requirements of astrometric performances.

\bibliographystyle{aa}
\bibliography{chips}

\end{document}